\documentclass[11pt]{article}
\usepackage{erice,epsfig}

\def\be{\begin{equation}}
\def\ee{\end{equation}}
\def\bea{\begin{eqnarray}}
\def\eea{\end{eqnarray}}

\begin{document}
\vspace*{4cm}
\title{Aspects of Coulomb Dissociation and Interference in Peripheral \\ Nucleus-Nucleus Collisions}

\author{Joakim Nystrand$^1$ \footnote{presenting author} , Anthony J. Baltz$^2$ and Spencer R. Klein$^3$}

\address{$^1$Department of Physics, Lund University, SE-221 00 Lund, Sweden \\
         $^2$Brookhaven National Laboratory, Upton, NY 11973, U.S.A. \\
         $^3$Lawrence Berkeley National Laboratory, Berkeley, CA 94720, U.S.A.}

\maketitle

\begin{center}
{\it Presented at the Workshop on Electromagnetic Probes of Fundamental Physics, Erice, Italy, 
16-21 October, 2001.}
\end{center}
\vspace{0.3cm}

\abstracts{
Coherent vector meson production in peripheral nucleus-nucleus collisions is discussed. 
These interactions may occur for impact parameters much larger than the sum of the 
nuclear radii. Since the vector meson production is always localized to one of the nuclei, 
the system acts as a two-source interferometer in the transverse plane. By tagging the 
outgoing nuclei for Coulomb dissociation it is possible to obtain a measure of the impact 
parameter and thus the source separation in the interferometer. This is of particular 
interest since the life-time of the vector mesons are generally much shorter than the 
impact parameters of the collisions. 
}

\section{Coherent Peripheral Nuclear Collisions}

This presentation will discuss some aspects of nucleus-nucleus collisions without 
physical overlap, i.e. collisions with impact parameters, $b$, larger than the sum of the 
nuclear radii, $R$, i.e. $b>2R$. Particles can be produced in these collisions through an 
interaction of the fields of the nuclei. The interactions can involve both the 
electromagnetic and nuclear fields, but because of the short range of the nuclear 
force, purely nuclear processes are suppressed for $b>2R$. If the momentum 
transfers from the nuclei are small enough ($Q< \hbar c /R$), the fields couple 
coherently to all nucleons. This enhances the cross sections and gives the events 
an unique signature, which can be used for identification. 
The restrictions on the momentum transfer do not prevent the production of
heavy systems, however, in high-energy collisions. States with masses up to about 
$2 \gamma \hbar c/R$, where $\gamma$ is the Lorentz factor of the beams in the 
laboratory system, can be produced in a heavy-ion collider; this corresponds to 
masses of a few GeV/c$^2$ at the Relativistic Heavy-Ion Collider (RHIC) at Brookhaven 
National Laboratory and about 100~GeV/c$^2$ at the Large Hadron Collider (LHC) at
CERN. 

The electromagnetic field of a relativistic nucleus can be treated as an equivalent
flux of photons. This was first realized by Fermi~\cite{Fermi}, and the idea was 
further developed by Weizs{\"a}cker and Williams.~\cite{WW} The method is now 
known as the Weizs{\"a}cker-Williams method. For relativistic nuclei, the impact 
parameter is a well-defined variable, and the photon flux should be 
evaluated in impact parameter space.~\cite{CahnJackson} The density of photons  
with energy $k$ at a perpendicular distance $b$ ($b > R$) from the center of 
a nucleus is (in units where $\hbar = c = 1$)~\cite{Jackson}
\begin{equation}
   n(k,b) =
   \frac{dN_{\gamma}}{d k d^2b} = 
   \frac{\alpha Z^{2}}{\pi^2} \frac{1}{k b^2} \left( \frac{k b}{\gamma} \right)^2  
   K_1^2(\frac{k b}{\gamma}) \; .
\label{WWeq}
\end{equation}
Here, $\alpha$ is the fine structure constant, Z the charge of the ion, and 
$K_1$ a modified Bessel function. The total number of photons with energy $k$, 
$n( k )$, is obtained by integrating the photon density over all impact parameters 
for which there are no hadronic interactions (roughly $b > 2R$). 

The photons from one of the nuclei may interact with the other nucleus either 
electromagnetically or hadronically. Electromagnetic interactions include collective 
excitation of the nucleus into a Giant Dipole Resonance (GDR) and particle production  
through two-photon interactions. The hadronic interactions are usually divided the following
3 categories: Vector meson dominance, direct, and anomalous.~\cite{Torbjorn} The latter 
two involve interactions with a parton in the nucleus and do not lead to a rapidity gap 
between the produced particle and the target nucleus. The coherent photonuclear interactions 
that will be considered here will be treated in the framework of the 
vector meson dominance model. 

According to vector meson dominance, the photon interacts hadronically by first fluctuating into 
a vector meson, or more generally a $q \overline{q}$-pair, carrying the same quantum numbers 
(J$^{PC}$=1$^{- -}$) as the photon. The scattering amplitude for the photonuclear interaction 
is  
\begin{equation}
\frac{d \sigma}{dt} (\gamma A) = 
\sum_{V} \frac{4 \pi \alpha}{f_V^2} \frac{d \sigma}{dt} (VA) \; ,
\end{equation}
where $f_V$ is the photon vector meson coupling, and $\sqrt{t}$ is the 
momentum transfer from the target nucleus. The sum is over all applicable 
vector meson states. 

The photon vector meson couplings are constrained from data on the semi-leptonic 
decay widths, $\Gamma_{V \rightarrow e^+ e^-}$,
\begin{equation}
\label{coupling}
   \frac{f_V^2}{4 \pi} = \frac{1}{3} 
                         \frac{M_V \alpha^2}{\Gamma_{V \rightarrow e^+ e^-}} 
   \; ,
\end{equation}
where $M_V$ is the vector meson mass.~\cite{Donnachie}   

The coherent reaction $\gamma A \rightarrow V A$  thus corresponds to  
elastic scattering, $V A \rightarrow V A$ if the cross terms are neglected (i.e. 
cases where the photon fluctuates into a state V and then scatters off 
the target into a state V'). The nuclear momentum transfer can be 
treated as exchange of a meson or a Pomeron, and the vector meson is produced 
through $\gamma$-Pomeron (or $\gamma$-meson) fusion. The Pomeron is the colorless 
exchange particle of the strong force and carries the same quantum numbers as 
the vacuum (J$^{PC}$=0$^{+ +}$).

The nuclear transverse momentum transfer is determined by the nuclear form factor, 
$F(t)$, and the photonuclear cross section is 
\begin{equation}
   \sigma( \gamma A \rightarrow V A) = \left. \frac{d\sigma}{dt} \right|_{t=0} 
   \int_{t_{min}}^{\infty} \left| F(t) \right|^2 dt \; ,
\end{equation}
where $t_{min}=(M_V^2/4 \gamma k)^2$ is the minimum momentum transfer needed to 
produce a vector meson. The form factor vanishes for $t \gg (1/R)^2$ and the 
cross section will thus go to zero for small photon energies $k < M_V^2 R/4 \gamma$. 

Convoluting the photonuclear cross section with the photon spectrum gives the total 
cross section for vector meson production in nucleus-nucleus collisions:
\begin{equation}
\label{convolution}
   \sigma(A+A \rightarrow A+A+V) = 2 \int_{0}^{\infty} n(k) \sigma_{\gamma A \rightarrow V A} (k) d k
   \; .
\end{equation}
The integral goes from $0$ to $\infty$; in practice it is cut-off at high $k$ by 
the exponential fall-off in the photon spectrum for $k > \gamma / R$ and at low 
$k$ by the nuclear form factor of the target. The factor of 2 takes into account 
that the nuclei can act both as target and photon emitter. 

The reaction $\gamma p \rightarrow V p$ has been studied at HERA and in fixed target 
experiments.~\cite{Crittenden} The cross sections in nucleus-nucleus collisions have been 
calculated in,~\cite{PRC} using the Weizs{\"a}cker--Williams photon spectrum and a Glauber 
for the photonuclear cross sections. The Glauber model calculations use the experimentally 
determined photon-proton cross sections as input. 
The results for gold at RHIC and lead at the LHC are given in 
Table~1. The cross sections are very large, roughly 10\% and 50\% of the hadronic cross 
section for $\rho^0$ production with gold at RHIC and lead at LHC, respectively. 

\begin{table}[t]
\caption{Cross sections for exclusive vector meson production at RHIC ($\gamma = $108) 
and at the LHC ($\gamma = $2940).}
\vspace{0.4cm}
\begin{center}
\begin{tabular}{ccc}
\hline
Vector Meson & \multicolumn{2}{c}{ $\sigma$ [mb] } \\
             & RHIC Au+Au &  LHC Pb+Pb \\ \hline
$\rho^0$     & 590        &  5200 \\ 
$\omega$     &  59        &  490  \\
$\phi$       &  39        &  460  \\
$J / \Psi$   &  0.29      &  32   \\ \hline
\end{tabular}
\end{center}
\end{table}

\begin{figure}[htb]
    \epsfxsize=0.5\textwidth
    \centerline{\epsffile{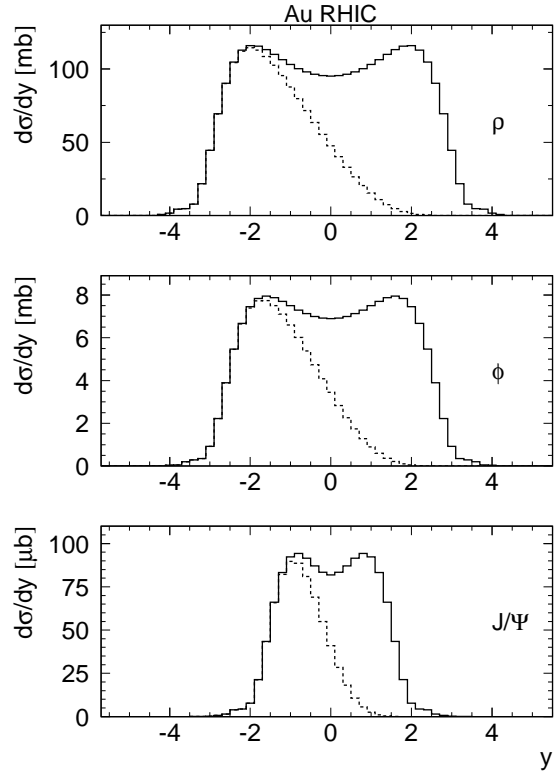}} 
    \label{fig1}
    \caption{Rapidity distributions of vector mesons in Au+Au interactions at RHIC. The dashed histograms show
the rapidity distributions when the photon is emitted by the nucleus with positive rapidity}
\end{figure}

The production is centered around mid-rapidity with a width determined by the mass of the vector meson. 
The rapidity, $y$, is related to the photon energy, $k$, through
\begin{equation}
   y = \ln( \frac{2 k}{M_V} ) \; .
\end{equation}
The shape of the rapidity distribution is determined by the photon spectrum and the 
energy variation of the photonuclear cross section and hence by the nuclear form factor. 
Three examples of vector meson rapidity distributions in Au+Au interactions at RHIC are 
shown in Fig.~\ref{fig1}. Further details about the vector meson production calculations 
are given elsewhere.~\cite{PRC}

\section{A two-source interferometer}

The two nuclei can act both as target and photon emitter in 
nucleus-nucleus collisions. This is different from $eA$ collisions where the electron  
emits the photon and the vector meson production is localized to the target nucleus. The 
transverse plane in a nuclear collision at impact parameter b is illustrated in Fig.~2. 

The impact parameter is not measured in the interactions. The cross section is obtained by 
integrating over all possible impact parameters, $b > 2R$, as was discussed in the previous section. 
The impact parameter dependence comes from the photon spectrum (Eq.~\ref{WWeq}). The calculations 
in the previous section assumed 
that the total cross section is given by the sum of the cross sections for production from the 
two sources (Eq.~\ref{convolution}). This is a reasonable assumption in most cases. For transverse momenta,
$p_T$, of the produced meson smaller than $1/b$, it is however not possible to distinguish which nucleus 
the meson came from, as can be understood from Fig. 2. One then has to add the amplitudes.~\cite{PRL}  

\begin{figure}[htb]
    \epsfxsize=0.55\textwidth
    \centerline{\epsffile{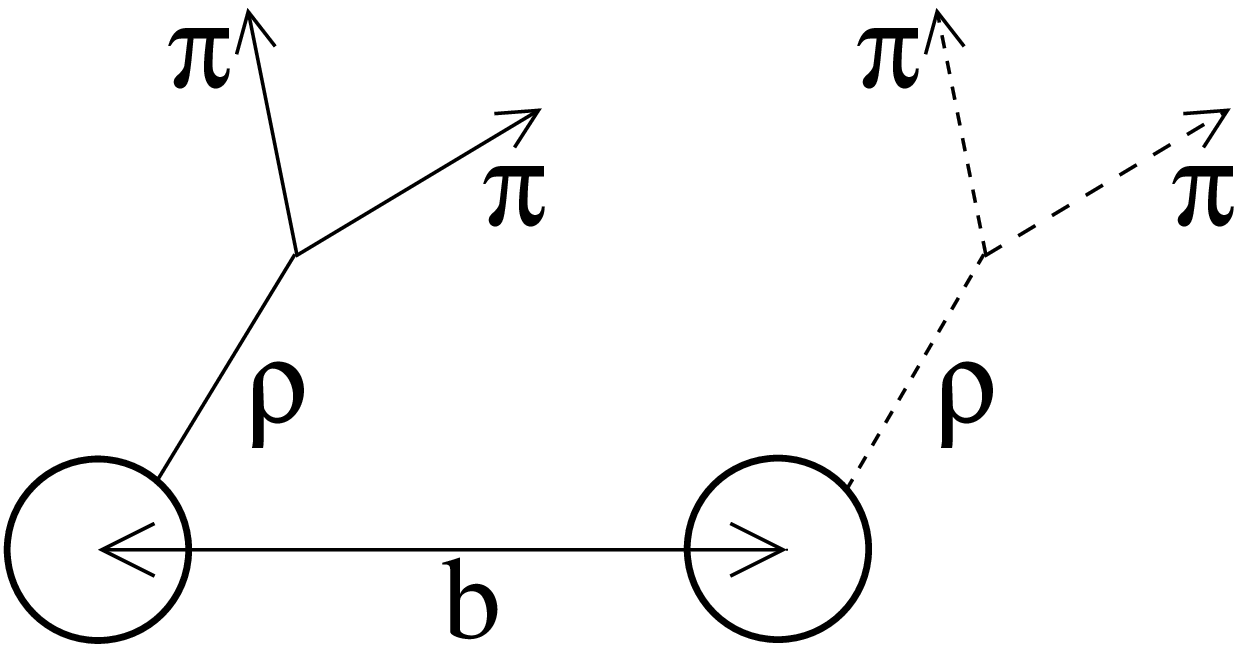}} 
    \label{fig2}
    \caption{The transverse plane for $\rho^0$ production. The production is always localized to one of the
nuclei because of the short range of the nuclear force. When the transverse momentum of the $\rho^0$-meson 
is smaller than $1/b$, it is not possible to distinguish at which nucleus it was produced.}
\end{figure}

The $p_T$ of the vector meson is the sum of the photon and Pomeron transverse momenta. 
The coherent couplings of the fields ensures that the $p_T$ will be of the order of $p_T < 1/R$.  
The transverse momentum distribution is given by the convolution of the photon, $f_1$, and Pomeron, $f_2$, 
distributions, 
\begin{equation}
\frac{dn}{d p_T} = f_{1,2}(p_T) = \int f_1 ( \vec{p}_T' ) f_2 ( \vec{p}_T - \vec{p}_T' ) 
d^2 \vec{p}_T' .
\end{equation}

The transverse momentum distribution of virtual photons of energy $k$ 
is given by \cite{gammapt}
\begin{equation}
   \frac{dN_{\gamma}}{d k_{\perp}} \propto 
   \frac{ | F( (k/\gamma)^2 + k_{\perp}^2 ) |^2 }{ ( (k/\gamma)^2 + k_{\perp}^2)^2 } k_{\perp}^3 .
\end{equation}
Similarly, the distribution of the nuclear transverse momentum transfer, 
$q_{\perp}$, is determined by the form factor \cite{PRL}
\begin{equation}
   \frac{d \sigma(\gamma A)}{d q_{\perp}} \propto  
   | F( t_{min} + q_{\perp}^2 ) |^2 \, q_{\perp} .
\end{equation}
The transverse momentum distribution of the photon is generally narrower than that of the Pomeron, 
so the vector meson $p_T$ distribution is dominated by the nuclear form factor. 

The differential cross section, $d \sigma / dy dp_T$, for a vector meson with 
rapidity, y, and transverse momentum, $p_T$, is the product of the photon density
and the photonuclear cross section, i.e. 
\begin{equation}
\label{intint}
\frac{d \sigma}{dy dp_T} = \int_{b>2R} 
k_1 \frac{dN}{d k_1 d^2b} \, \sigma(\gamma A_2) \, f_{1,2}(p_T) 
\; + \; 
k_2 \frac{dN}{d k_2 d^2b} \, \sigma(\gamma A_1) \, f_{2,1}(p_T) 
\; d^2 \vec{b}   \; .
\end{equation}
This can be written
\begin{equation}
   \label{noint}
   \frac{d \sigma}{dy dp_T} = \int_{b>2R} \left( \left| A_1 \right|^2  + \left| A_2 \right|^2 \right) d^2 \vec{b} 
   \; ,
\end{equation}
where the $A_1$ and $A_2$ correspond to the amplitudes for production off each of the two nuclei. 
This is only valid for $b \gg 1 / p_T$. The appropriate form for all
impact parameters is instead
\begin{equation}
   \label{interference}
   \frac{d \sigma}{dy dp_T} = \int_{b>2R} \left| A_1  +  A_2 \right|^2  d^2 \vec{b} \; .
\end{equation}
At mid-rapidity, the amplitudes will be of equal magnitude because of symmetry, and 
by comparing Eq.~\ref{intint} with Eq.~\ref{noint}, one sees that
\begin{equation}
\left| A_1 \right|^2 = \left| A_2 \right|^2 = 
k_1 \frac{dN}{d k_1 d^2b} \, \sigma(\gamma A_2) \, f_{1,2}(p_T) \; .
\end{equation}

Production at mid-rapidity is the only case that will be considered here; 
the general case $y \neq 0$ is discussed elsewhere.~\cite{PRL} To properly calculate the differential 
cross section from Eq.~\ref{interference}, one has to determine the relative phases of $A_1$ and $A_2$. The 
vector meson has negative parity so $A_1$ and $A_2$ have opposite signs; exchanging the 
positions of the two nuclei is equivalent to a reflection of the spatial coordinates, 
$\vec{x} \rightarrow - \vec{x}$. 

To a distant observer at the position $\vec{r}$, a vector meson produced at the position of 
nucleus~1, $\vec{x}_1$, will have a different phase from one produced at the position of 
nucleus~2, $\vec{x}_2$, because of the different path lengths. It is a reasonable assumption here, 
and it simplifies the calculations considerably, that the emission is from two point sources. The 
size of the nuclei is thus neglected. This is justified, since the nuclear dimensions 
($\sim$~7~fm) are much smaller than the typical impact parameters. It is furthermore assumed that 
the vector mesons can be treated as plane waves. The amplitude for a single source is then 
\begin{equation}
A_i = A_0 \cdot e^{ i \vec{p} \cdot \vec{x}_i } \; .
\end{equation}
where $A_0$ is the magnitude of $A_i$. The sum of the amplitudes is then 
\begin{equation}
A_1 + A_2 = 2 A_0 \, exp( \frac{i}{2} \left[ \vec{p} \cdot ( \vec{x}_1 + \vec{x}_2 ) + \pi \right] )
            \, \sin( \frac{ \vec{p} \cdot \vec{b} }{2} )
\end{equation}
and the magnitude squared is
\begin{equation}
\label{costerm}
\left| A_1 + A_2 \right|^2 = 2 A_0^2 \left( 1 - \cos( \vec{p} \cdot \vec{b} ) \right) \; .
\end{equation}
The impact parameter is $\vec{b} = \vec{x}_1 - \vec{x}_2$. This is the same interference pattern as 
from a two-source interferometer with slit separation $| \vec{b} |$, albeit with the opposite sign 
(destructive interference). 

This system is 
particularly interesting because the vector mesons, except for the $J/ \Psi$, are 
short-lived in comparison with the source separation divided by the speed of light; the life-times are 
$\rho^0$ 1.3 fm/c, $\omega$ 23 fm/c, $\phi$ 44 fm/c, and $J / \Psi$ 2300 fm/c. The interference will thus 
involve the decay products, which are in an entangled state, rather than the vector meson itself. 
Observing the interference would thus be a 
proof of entanglement and hence an example of the Einstein-Podolsky-Rosen paradox.~\cite{PRL} 

Using Eq.~\ref{costerm}, the integral in Eq.~\ref{interference} can be written
\begin{equation}
   \frac{d \sigma}{dy dp_T} = 2 \int_{b>2R} A_0^2 \left( 1 - \cos( \vec{p} \cdot \vec{b} ) \right) \; .
\end{equation}
For large values of $p_T$, the term containing $\cos( \vec{p} \cdot \vec{b} )$ will go through several
oscillations in the integration over b, and the net contribution to the integral will go to zero. For 
small transverse momenta however, $p_T \ll 1 / \langle b \rangle$, $\vec{p} \cdot \vec{b} \approx 0$ for all relevant 
impact parameters, and the interference term will lead to a vanishing cross section in this region. 
Since the median impact parameters for light vector mesons are around 40~fm at RHIC and about 
200~fm at the LHC, one expects the interference to be significant for $p_T < 5$~MeV/c at RHIC and 
$p_T < 1$~MeV/c at the LHC. 

The transverse momentum distribution calculated with and without interference is shown in Fig.~4, but before 
the results are discussed further, the consequences of vector meson production in coincidence with nuclear 
Coulomb break-up will be considered.

\section{Electromagnetic dissociation}

The photon spectrum (Eq.~\ref{WWeq}) is proportional to $Z^2$ and inversely proportional to $k$ and $b^2$. 
The density of low-energy photons is thus high for small b. 
This may lead to violation of unitarity, as can be seen by rewriting the cross section as an interaction probability 
for a given impact parameter,  
\begin{equation}
   \label{amplitude}
   P^1 (b) = \frac{d ^2 \sigma}{db^2} = 
             \int \frac{dN_{\gamma}}{d k d^2b} \sigma_{\gamma A \rightarrow X} (k) d k \; .
\end{equation}
The superscript '1' on $P$ indicates that it is the first order probability. For photonuclear 
processes with low thresholds and/or high cross sections $ P^1 (b)$ may exceed one 
for small impact parameters. 

This is the case for a general photonuclear excitation followed by nuclear break-up, 
$A + B \rightarrow A + \gamma + B \rightarrow X + B$, in gold and lead interactions at 
RHIC and the LHC. Clearly, Eq.~\ref{amplitude} cannot be interpreted as a probability under such 
circumstances. Instead, $P^1 (b)$ should, in the general case, be interpreted as a first-order 
amplitude. Unitarity can then be restored by accounting for multiple excitations, where the probability of 
having exactly $N$ excitations is given by a Poisson distribution, 
\begin{equation}
   P_N (b) = \frac{ \left( P^1 (b) \right)^N \, e^{- P^1 (b)} }{N!} \; .
\end{equation}

The cross section for a photonuclear excitation of a single nucleus in Au+Au interactions at RHIC 
is very large, about 95~barns.~\cite{BRW} The largest contribution to this (65~barns) is excitation 
of the target nucleus into a Giant Dipole Resonance (GDR). The GDR decays by emitting a single neutron 
in about 83\% of the events.~\cite{veyssiere} 
 
The cross section for mutual Coulomb excitation of both ions was calculated in,~\cite{BCW} assuming 
that the two excitations are independent. Two cases were treated: a general excitation into 
a state which results in the emission of an arbitrary number of neutrons (Xn), and an excitation into 
a GDR which results in the emission of a single neutron (1n). The total cross section for mutual excitation
was found to be 11~barns and of this roughly 7~barns are from hadronic interactions and about 4~barns from 
Coulomb dissociation. The cross section for mutual GDR excitation followed by the emission of exactly one 
neutron in both directions was calculated to be 0.4~barn. 

This Coulomb cross section is of importance for the trigger in experiments at RHIC and the LHC. All RHIC 
experiments are equipped with identical Zero-Degree Calorimeters (ZDCs) for triggering and event selection. 
The calorimeters, situated on both sides of the experiments approximately 18~m from the center of the 
interaction region, detect free neutrons in the forward direction from the fragmentation of the beam nuclei. 
Calculations 
indicate that about 1/3 of the ZDC coincidence triggers are from Coulomb interactions. 
Results from the first year of data taking at RHIC are in good agreement with the calculations,~\cite{ZDCdata} 
which indicates that the assumption of independent excitations is a reasonable one. 

These large cross sections, and hence probabilities, suggest that it should be possible to 
observe vector meson production in coincidence with mutual dissociation. This has indeed been 
observed by the STAR collaboration.~\cite{STAR} If the Coulomb break-up and the 
vector meson production are independent, the probabilities factorize, and the cross sections can be 
calculated along the same lines as for mutual Coulomb excitation. The validity of the assumption 
of factorization is hard to prove rigorously. It is reasonable if one considers the different time 
scales involved. The lifetime of a GDR in a heavy nucleus is about 50~fm/c in the rest 
frame of the nucleus,~\cite{GDR} whereas the typical hadronic formation time is of the order of 1~fm/c
in the center of mass.~\cite{Bjorken} 
The nucleus, if it is excited into a GDR, will thus remain intact during and long after the 
production of the vector meson. The fact that the protons and neutrons are in a state of collective 
oscillations is not expected to affect the coherent vector meson nucleus scattering significantly. 

Vector meson production in coincidence with Coulomb excitation leading to a final state with 
(Xn,Xn) and (1n,1n) neutrons will be considered. The probability for mutual Coulomb excitation and no 
hadronic interaction at an impact parameter b is
\begin{equation}
   P_{C} (b) = \left( 1 - 2 e^{ - P_{C}^1 (b) } +  e^{ - 2 P_{C}^1 (b) } \right)  e^{ - P_H^1 (b) }
\end{equation} 
where $P_{C}^1 (b)$ is calculated from Eq.~\ref{amplitude} with $\sigma (\gamma A)$, 
the total photonuclear cross section, 
taken from a parameterization 
of experimental data.~\cite{BCW} $P_H^1 (b)$ is the first-order hadronic interaction probability, 
which is calculated from the Glauber model
\begin{equation}
P_H^1 (b) = \int T_A (\vec{r} - \vec{b}) \left( 1 -  \exp( - \sigma_{NN} T_B(\vec{r}) \right) d^2\vec{r} ) \; ,
\end{equation}
where $\sigma_{NN}$ is the total nucleon-nucleon cross section and $T_{A,B}$ is the nuclear thickness 
function.~\cite{Glauber} $P_{C} (b)$ has a maximum of about 35\% at $b = 15$~fm in Au+Au interactions
at RHIC. 

Similarly, the probability for mutual Coulomb excitation leading to the emission of single neutrons is
\begin{equation}
   P_{C(1n,1n)} (b) = \left( P_{C(1n)}^1 (b) \right)^2 \, e^{ - 2 P_{C}^1 (b) - P_H (b) }
\end{equation} 
where $P_{C(1n)}^1 (b)$ is also calculated from Eq.~\ref{amplitude} but with $\sigma(\gamma A)$ 
being the cross section for GDR excitation followed by single neutron emission.

\begin{table}[t]
\caption{Cross sections for exclusive vector meson production in Au+Au at RHIC ($\gamma = $108) 
and Pb+Pb at the LHC ($\gamma = $2940) with nuclear Coulomb dissociation.}
\vspace{0.4cm}
\begin{center}
\begin{tabular}{ccc|cc}
\hline
Vector Meson & \multicolumn{4}{c}{ $\sigma$ [mb] } \\
             & \multicolumn{2}{c}{RHIC Au+Au}    &  \multicolumn{2}{c}{LHC Pb+Pb} \\ 
             & 1n,1n      & Xn,Xn       & 1n,1n        &  Xn,Xn   \\ \hline
$\rho^0$     & 3.7        & 42          & 12           &  210     \\ 
$\omega$     & 0.34       & 3.9         & 1.1          &  19       \\
$\phi$       & 0.27       & 3.1         & 1.1          &  20       \\
$J / \Psi$   & 0.0036     & 0.044       & 0.14         &  2.5     \\ \hline
\end{tabular}
\end{center}
\end{table}

The distribution of impact parameters will thus be altered in interactions where the nuclei dissociate.  
The integral over b (Eq.~\ref{interference}) has to be weighted with the 
Coulomb probabilities $P_{C} (b)$ or $P_{C(1n,1n)} (b)$: 
\begin{equation}
   \frac{d \sigma(Xn,Xn,V)}{dy dp_T} = \int P_{C} (b) \, \left| A_1  +  A_2 \right|^2  d^2 \vec{b} \; . 
\end{equation}

The total vector meson cross sections, integrated over $y$ and $p_T$, at RHIC and the LHC are presented in Table~2. 
Requiring coincidence with a general photonuclear interaction (Xn,Xn) reduces the cross sections with roughly 
a factor of 10 at RHIC and up to almost a factor of 30 at the LHC. The cross section with emission of a single 
neutron in each direction is about 10\% of the cross section in coincidence with a general photonuclear interaction.

\begin{figure}[htb]
    \epsfxsize=0.7\textwidth
    \centerline{\epsffile{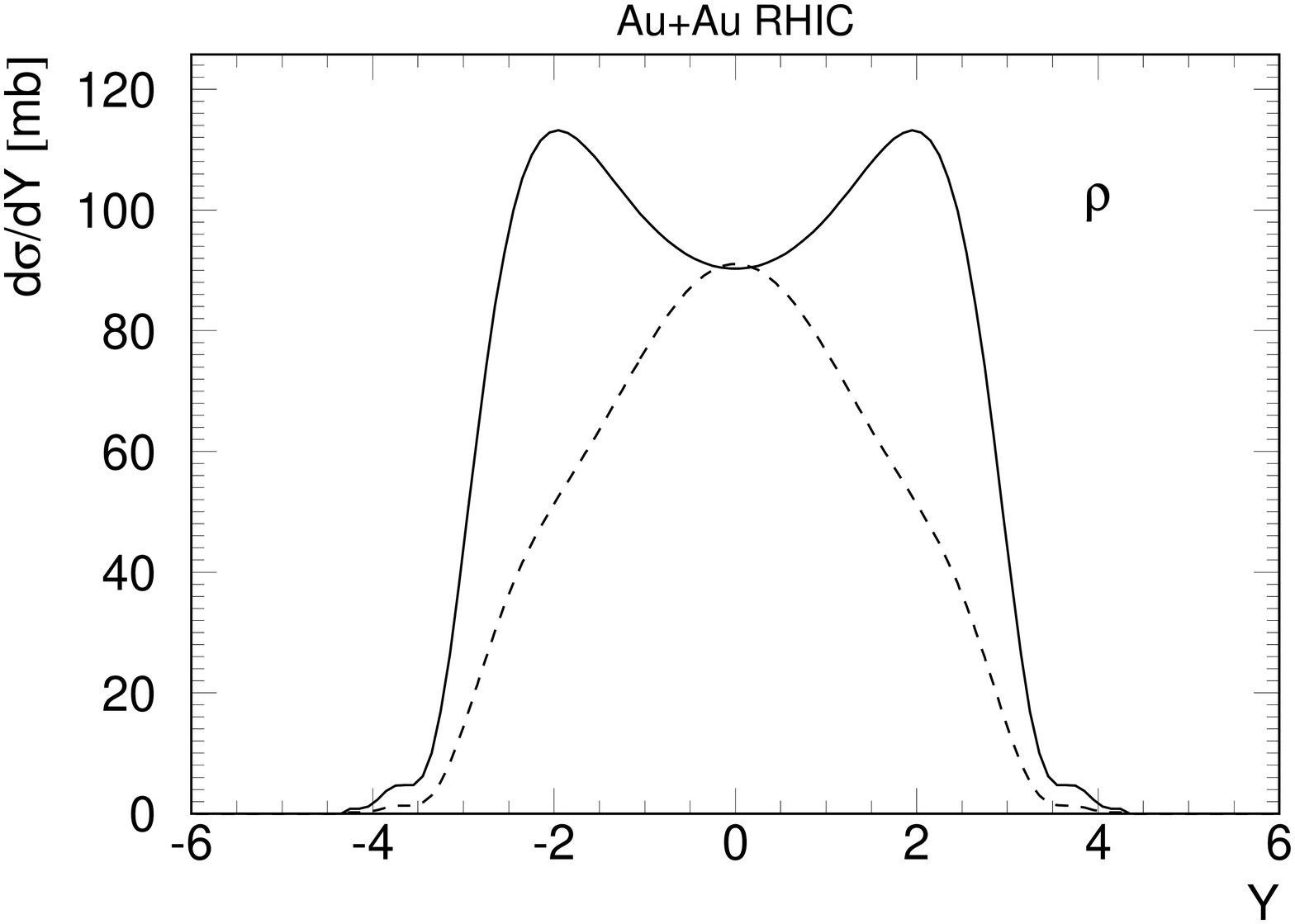}} 
    \label{fig3}
    \caption{Rapidity distributions of $\rho^0$-mesons in Au+Au interactions at RHIC with (dashed curve) and 
without (solid curve) nuclear breakup. The curve with breakup has been multiplied by a factor of 10.}
\end{figure}

The modified impact parameter distribution also affects the photon spectrum, as can be understood from 
Eq.~\ref{intint}. The increased relative abundance of small impact parameters in 
interactions with breakup hardens the 
photon spectrum. This modification of the photon spectrum leads to a corresponding change in the 
rapidity distribution, favouring production at central rapidities. This is 
illustrated in Fig.~3, where the rapidity distribution of $\rho^0$ mesons is calculated with and 
without break-up.

\begin{figure}[htb]
    \epsfxsize=0.7\textwidth
    \centerline{\epsffile{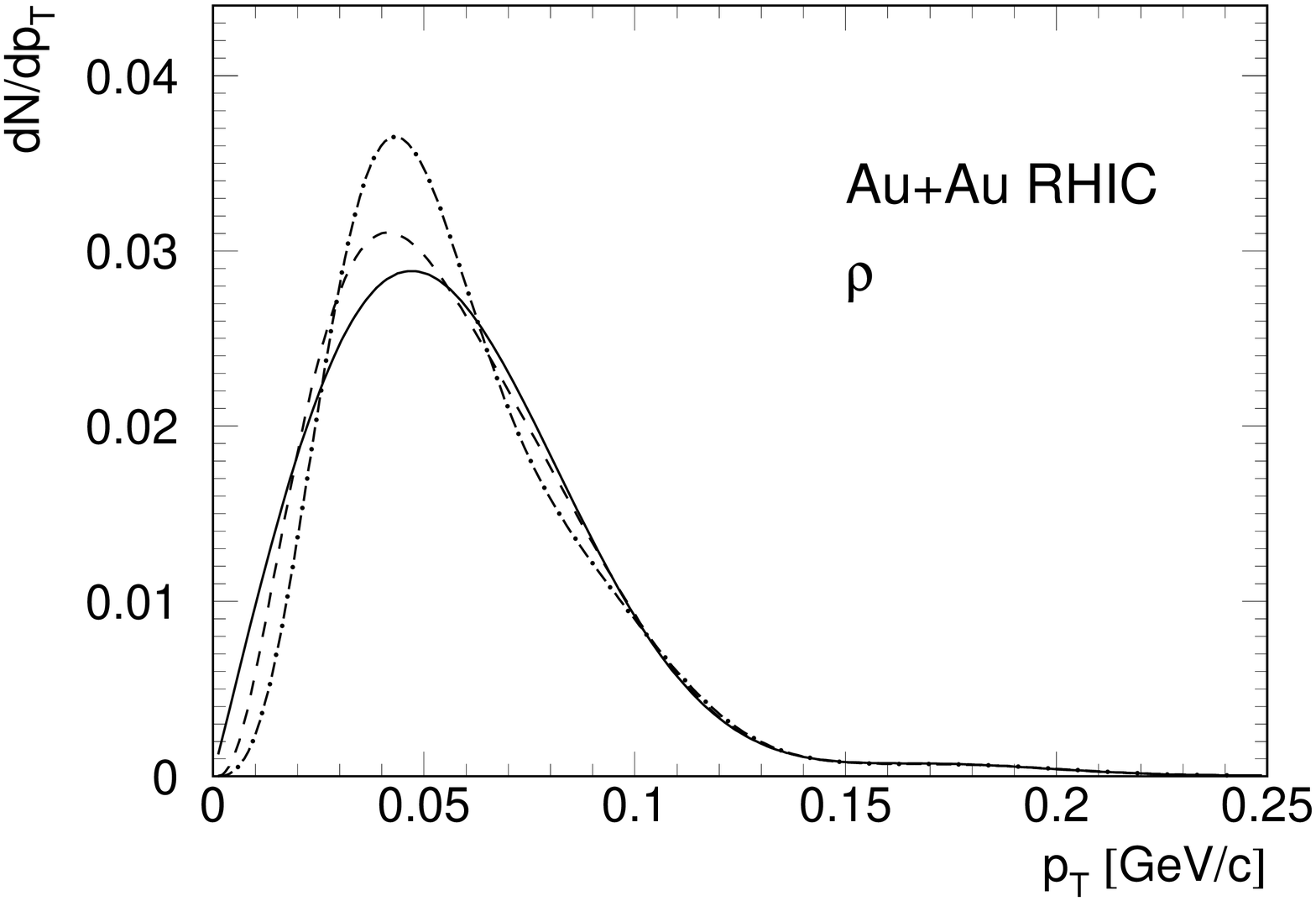}} 
    \label{fig4}
    \caption{Transverse momentum distributions of $\rho^0$ mesons in Au+Au interactions at RHIC. The solid curve is 
without interference, the dashed with interference in exclusive production, the dot-dashed with 
interference and Coulomb break-up (Xn,Xn).}
\end{figure}

The altered impact parameter distribution is, furthermore, of importance for the transverse momentum   
spectrum and the interference discussed previously. The two nuclei act as a two-source 
interferometer in the transverse plane. By requiring mutual Coulomb dissociation of the nuclei it is possible 
to obtain a handle on the source separation. The median impact parameters are 
approximately a factor of two smaller in interactions with Coulomb breakup compared with exclusive 
production. The interference is thus expected to be significant at correspondingly higher transverse momenta. 
The calculated $p_T$ spectrum for $\rho^0$ mesons produced in Au+Au interactions at RHIC is shown in Fig.~4 for 
interference in exclusive production, and interference in reactions with Coulomb dissociation (and, for comparison, 
without interference). 
The plot shows $dN/dp_T$, as opposed to $dN/dp_T^2$, so all curves go to zero at low $p_T$ because 
of vanishing phase space. As expected, the interference is stronger and reaches larger $p_T$ in interactions 
with Coulomb breakup.

\section{Conclusions}

The consequences of vector meson production in coincidence with nuclear Coulomb dissociation in peripheral 
heavy-ion interactions at RHIC and the LHC have been studied. Production in coincidence with nuclear 
breakup alters the phase space distribution of the vector mesons compared with exclusive production. 
The breakup requirement enables one to get an independent handle of the impact parameter of the 
interactions. This will benefit the understanding of the predicted interference phenomenon.

\section*{Acknowledgments}
This work was supported by the Swedish Research Council (VR).


\end{document}